# Demonstration of plasma focus operation without using sliding discharge on a glass or ceramic insulator for plasma initiation

S K H Auluck and Srushti Parhar
HiQ TechnoWorks Pvt Ltd, Navi Mumbai, India

M. G. Kulkarni, Anil Magtum, Suresh Ghatge, Nikita Suryavanshi and P. U Mirajkar
Rectiphase Capacitors Pvt. Ltd, Nashik, India

**Abstract**

This paper reports on experimental work related to a technology development project that had to be abandoned in the wake of the COVID-19 pandemic. Recognizing that the window of opportunity for timely exploitation of this technology is no longer open and also taking into account the likely future commercial value of the downstream innovations, it was decided to partially release proprietary experimental information which is of current scientific interest to the plasma focus community. This information pertains to the replacement of the traditional plasma initiation method using a sliding discharge on a glass or ceramic insulator with a novel construction. This paper discusses the motivation, theoretical background, practical realization and experimental observations. Some downstream novel plasma focus based technology concepts that are heavily dependent on this innovation are briefly described.

## I. Introduction:

Plasma focus is a relatively simple and inexpensive plasma device[1-4] that produces many effects which could be exploited to generate commercial value. Material processing and nanotechnology [5], production of short-lived radioisotopes for nuclear medicine [6], intense pulsed neutron sources for detection of concealed containerized organic contraband (explosives, narcotics, human trafficking) [7,8] are some of the avenues for such commercial developments. Such efforts usually do not make much progress in the commercial realm because of a fundamental lack of control on the process of initial plasma formation, which is based on generation of a sliding plasma discharge on a suitably chosen insulator material. Experimental reports [9-15] reveal the tentative nature of understanding of this process and currently available numerical modeling [16-18] reports do not show any signs of pointing the way towards exercising any optimizing control either.

The authors embarked on a technology development project based on the plasma focus that had as its core, a potential solution to the plasma initiation problem and a pathway towards significant scaling up. This experimental program of the project was delayed by a series of force majeure events (including a medical setback) that culminated in the COVID-19 pandemic and had to be ultimately abandoned. Recognizing that the window of opportunity for timely exploitation of this technology is no longer open and also taking into account the likely future commercial value of the downstream innovations, it was decided to partially release proprietary experimental information which is of current scientific interest to the plasma focus community. This is the background and motivation of the present paper.

The paper is organized as follows. Section II provides a conceptual model for the proposed solution and its experimental realization in a proof-of-concept configuration. Section III describes experimental results demonstrating occurrence of plasma focus action. Section IV discusses promising downstream technologies. Section V summarizes and concludes the paper.

II. <u>Conceptual model for the proposed solution and its realization:</u>

The problem of formation of the initial plasma in the plasma focus can be restated as uniform gaseous ionization in a thin annular cylindrical space that allows the resulting plasma to separate from the space-defining structure and travel away from that space. The breakdown process must necessarily occur as a travelling ionization wave over a guiding surface rather than as simultaneous ionization of all the gas in the designated volume since the latter process cannot prevent ionization from spilling over outside the designated volume.

Gaseous breakdown in the extended free space between two electrodes is usually non-uniform because of formation of streamers that have smaller transverse dimensions than their length in the direction of propagation [19]. They also have a tendency to formation of branches in a fractal pattern. One can be certain of uniform discharge if the size of the volume is comparable to the size of a streamer. Our proposed solution to the problem of launching an ionization wave over a cylindrical surface is explained with respect to Fig 1.

This is a schematic of a ladder circuit that involves identical self-breaking spark gaps in series connection. Each spark gap has a small stray capacitance $C_G$ across it and each junction between adjacent spark gaps is connected to a common point with a capacitance $C_0$. When a high voltage pulse of amplitude $V_0$ is applied between the free end of the ladder and the common point, a capacitive division occurs and the voltage $V_1$ appears across the first spark gap.

$$V_1 = \frac{C_0}{C_0 + C_G} V_0 \tag{1}$$

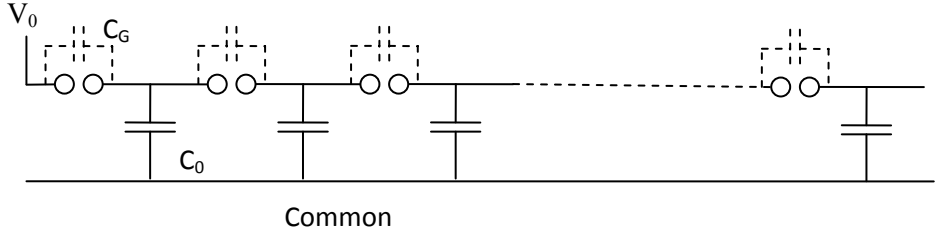

Fig 1: Schematic of a ladder circuit that results in sequential self-breakdown of each spark gap

This is much larger than the self-breakdown voltage of the first spark gap so the gap fires and the voltage $V_1$ appears at the junction between the first and the second spark gap. Again this is divided between the stray capacitance of the second spark gap and the capacitance $C_0$ connecting its junction with the next spark gap to the common so that a voltage

$$V_2 = \frac{C_0}{C_0 + C_G} V_1 = \left(\frac{C_0}{C_0 + C_G}\right)^2 V_0 \tag{2}$$

appears across the second spark gap. Continuing in this manner, it can be seen that the $n^{th}$ spark gap in the series will get the voltage

$$V_n = \left(\frac{C_0}{C_0 + C_G}\right)^n V_0 \tag{3}$$

When $C_G \ll C_0$, it is readily seen that the voltage decrease at each stage of the ladder is quite small. If $V_0$ is much higher than the breakdown voltage of each spark gap, an ionization wave propagates down the ladder.

This scheme is implemented in the manner shown in Fig. 2, that ensures field intensification across each spark gap that is known to decrease the time-to-breakdown. The "common" is the anode core that is covered with an insulator sleeve machined out of an ordinary plastic like polypropylene (see Fig 3). The cascade of spark gaps is made of a stack of alternating insulating and metallic rings. The metallic rings are made out of 50 μm thick stainless steel sheet by photochemical milling. The insulating washers are made out of 200 µm mylar sheet by laser cutting. Both have an outer diameter of 14 mm that matches the anode diameter and inner diameter that is a sliding fit to the insulating sleeve.

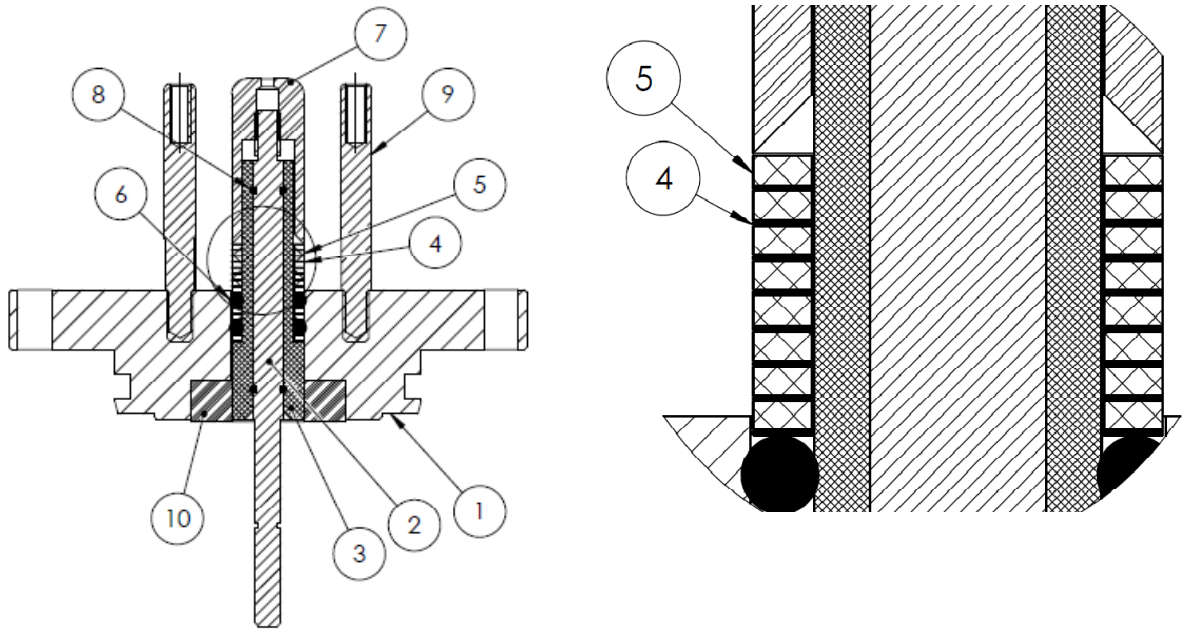

Fig 2. (1) PF support (2) Anode core (3) Insulating sleeve (4) metallic washers (5) Mylar washers (6) Outer O rings (7) Anode cap (8) Inner O rings (9) Cathode rods 6 mm diameter on 34 mm centers (8 Nos) (10) Silicone rubber gasket

The stack of washers is pressed against the step in the insulating sleeve by a stainless steel anode cap that is threaded over the anode core. A light coating of vacuum grease is applied over the insulating sleeve to prevent breakdown between the inner edges of the adjacent metal rings. The anode assembly has an overall diameter of 14 mm. The plasma focus assembly is electrically connected to the capacitor using a flexible strip transmission line whose insulation is pressed against the silicone rubber gasket using a mechanical arrangement that allows a low inductance high current connection to the anode core. Fig 3 shows the parts and the assembly of the anode.

Note that the "common" and capacitors $C_0$ are pulse-charged to a high voltage on firing the main spark gap powering the plasma focus. The ionization wave proceeds as the spark gaps progressively discharge the capacitors $C_0$ to the plasma focus cathode.

This was meant to be a proof-of-concept experiment, to see whether plasma focus action is possible using an ionization wave generated on the outer surface of the washer stack, that should be an inside out washer gun. The mylar washers can be replaced by sapphire rings in industrial applications. In our experiments two mylar washers were used between a pair of stainless steel washers.

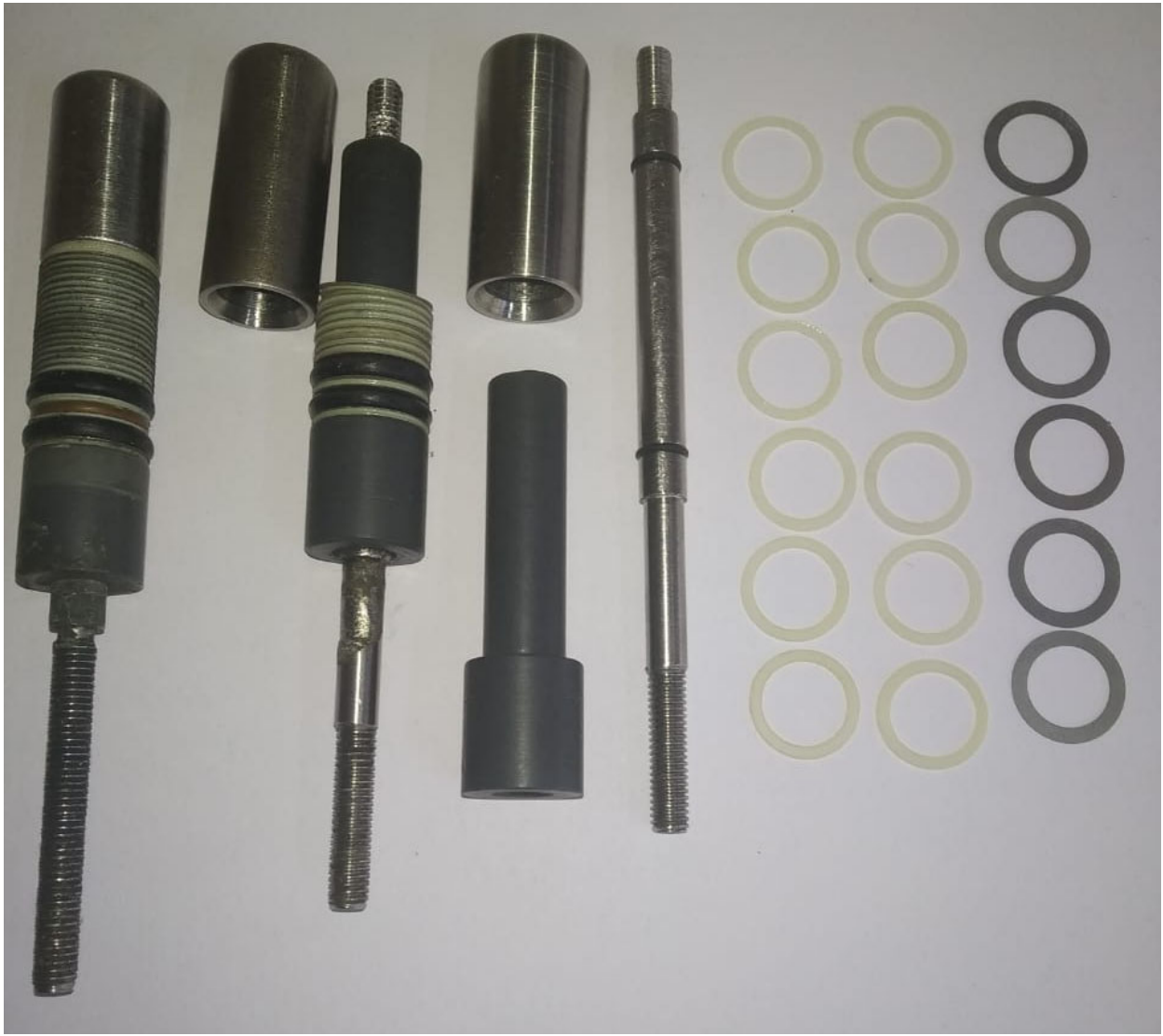

Fig 3. Photograph of the parts and the anode assembly.

III. **Experimental results**:

As mentioned above, the experimental program was delayed by a series of force majeure events. To save cost, a makeshift system was made for initial trials and based on the experience, improved hardware was designed and assembled. Unfortunately, we could not pursue our work with the improved hardware initially because of a capacitor leak followed by a medical setback and later because of the COVID 19 pandemic. We report here on the only noteworthy observation made with our makeshift hardware, namely, occurrence of the characteristic singularity in the current derivative waveform. This was observed with air as working gas. The pressure was 2-3 mbar and the voltage was 10-12 kV.

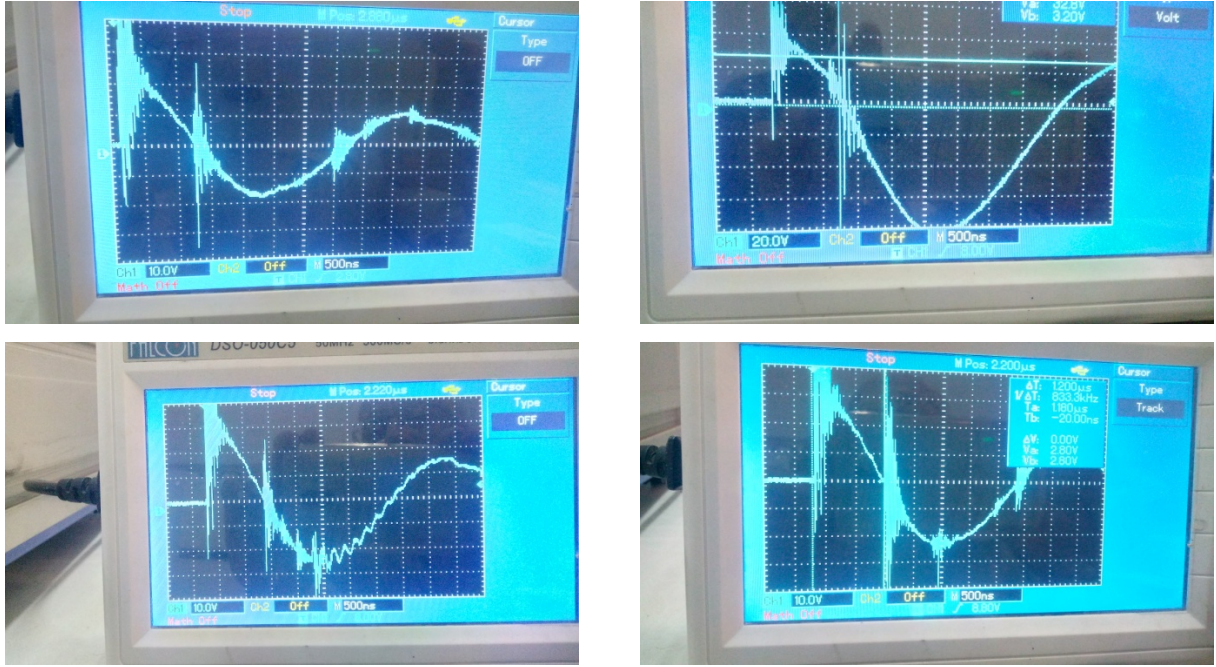

Fig 4: Observation of current derivative singularity near the maximum of current in air at few mbar pressure at 10-12 kV.

One of our observations was the necessity of about 20 training shots before the current derivative singularity made its appearance. This can be interpreted as initially non-uniform discharges between adjacent metal rings because of small scale non-uniformities on the edges of the metal rings that create localized zones of high electric field intensification. Repeated flow of high current would burn off those non-uniformities so that no preferred spots are left for discharge initiation. That ultimately leads to an axially uniform ionization wave over the surface of the stack. Impurities from the mylar washers subjected to intense plasma are expected to enter the plasma in this version. This problem should disappear when mylar is replaced with sapphire .

This construction removes one fundamental limitation of the plasma focus in commercial applications, namely, the tendency of the glass or ceramic insulators to break. Since no current flows through the metal washers after the plasma moves away, their erosion is minimal.

## IV. Significance of this work: downstream technologies

The significance of this invention is revealed below through a partial disclosure of proprietary technical information related to certain downstream technologies of commercial value

a) <u>High pressure optimized plasma focus devices as compact generators of nuclear reaction products</u>: It is clear that the atomic physics processes involved in the transition from neutral gas to partially ionized plasma are confined to the narrow annular region between the edges of adjacent metal rings at a time and not the entire volume that is ultimately traversed by the ionization wave. This circumstance should enable this transition to proceed rapidly even at higher gas pressures, making possible optimized plasma focus operation at higher than commonly used pressure – even ~1 bar of deuterium [20]. Such high pressure optimized plasma focus operation can be used for on-site breeding of short lived radio-isotopes for medical imaging using PET and as intense neutron source for boron-neutron brachytherapy for treatment of certain cancers.

b) <u>Plasma focus with in-built magnetic energy storage and switching</u>: Commercial applications of plasma focus are hampered by the necessity of a low inductance connection between the energy storage capacitor bank and the plasma focus device. This translates into an unwieldy interface that connects a large number of cables to the device, compromising its ability to be placed close to the user area. The present invention can avoid this by allowing placement of secondary magnetic storage and switching inside the anode. This is achieved by configuring the insulator sleeve that separates the washer stack from the anode using suitable ferrite rings that are epoxy-bonded into an integral piece capable of withstanding the operating high voltage. The initial application of voltage allows plasma initiation and detachment. During the initial current rise, the ferrite rings provide a high inductance so that the cable connections to the capacitor bank need not have a low inductance. The ferrite stack is designed to have the capability of magnetically storing a certain fraction of the energy in the capacitor bank. When this amount of energy is transferred into the device with an already formed plasma sheath, the ferrite saturates and drives the energy into the plasma sheath. A similar idea was implemented in the Nessi plasma focus where saturable ferrite rings were inserted over the cables to increase transit time isolation between the spark gaps without increasing the inductance of the discharge circuit at the pinch time.

c) <u>Nanosecond-duration repetitive intense neutron pulses for rapid non-intrusive detection of containerized concealed organic contraband</u>: A repetitive, serviceable, 14 MeV pulsed neutron source having a very short pulse duration, around $10^5 - 10^7$ neutrons per pulse, repetition rate of ~ 1−100 kHz can be developed by configuring the present innovation to work in vacuum, with the washers made from palladium or copper plated with palladium and loaded with deuterium. This works like an inside-out washer gun accelerating a small amount of deuterium desorbed from the washer stack in ultra-high vacuum. The plasma blob created at the end of the anode is then guided through a magnetic drift tube into an electrostatic accelerator, where a synchronized high voltage pulse ~ 150 kV accelerates the ions onto a continuously cooled tritium target.

d) <u>A large area plasma source for in-situ synthesis and annealing of nanostructures</u>: This extends the ideas proposed in Ref 5 to an industrial scale large area plasma source. Essentially, the present construction of the plasma focus is reduced in size to an anode radius of just 6 mm. Many such plasma foci are supported on a single support (equivalent to item 1 in Fig 2) in a hexagonal array. The cathode rods are placed at the vertices of the hexagonal lattice and the anodes are placed at the centers of the hexagons. The technique of making a low inductance high current connection between the plasma focus and a flexible high voltage strip transmission line is easily generalized to such an array. The multiple plasma focus discharges are self-synchronizing because if one plasma focus starts early, its inductance will increase faster than others and less current will flow through it slowing it down. The result is that post-pinch plasmas from the array will expand and merge at some distance from the anode faces. At such distance, the plasma environment is suitable for in-situ synthesis and annealing of nanostructures [5]. The difference lies in the much larger area covered by this array as compared with a single plasma focus. The "unbreakable" nature of the plasma initiation surface as against the brittle glass or ceramic insulators used in conventional plasma foci makes this a viable technology.

## V. <u>Summary and conclusion</u>

This paper discloses experimental work related to a technology development project that had to be abandoned in the wake of the COVID-19 pandemic. A construction involving a periodic placement of metal and insulator rings acts as a ladder circuit with many self-breaking spark gaps in series that fire sequentially mimicking the sliding discharge occurring on the glass or ceramic insulator in a conventional plasma focus. We present evidence of occurrence of the typical current derivative singularity indicating plasma focus action in air as working gas.

The significance of this invention is revealed through a partial disclosure of proprietary technical information related to certain downstream technologies of commercial value.

## VI. <u>References</u>